\documentclass[12pt]{article}

\usepackage[round, sort]{natbib}
\usepackage{amsmath, amssymb, amsthm, enumerate, paralist, framed}
\usepackage[usenames, dvipsnames]{color}
\usepackage{wrapfig, booktabs, adjustbox}
\usepackage{graphicx}
\usepackage[]{algorithm2e}
\usepackage[margin=1.5in]{geometry}
\usepackage{fancyhdr, lastpage}
\usepackage{pdfpages}
\usepackage{morefloats}
\usepackage{hyperref}
\usepackage{microtype}

\title{The Effect of Data Swapping on Analyses of American Community Survey
Data}
\author{Nicolas Kim\footnote{Department of Statistics, Carnegie Mellon
  University, Pittsburgh, PA, 
  \mbox{\href{mailto:nicolask@stat.cmu.edu}{nicolask@stat.cmu.edu}}.}
}
\date{}

\begin{document}
\maketitle

\begin{abstract}
Researchers from a growing range of fields and industries rely on public-access
census data. These data are altered by census-taking agencies to minimize the
risk of identification; one such disclosure avoidance measure is the data
swapping procedure. I study the effects of data swapping on contingency tables
using a dummy dataset, public-use American Community Survey (ACS) data, and
restricted-use ACS data accessed within the U.S.\ Census Bureau. These
simulations demonstrate that as the rate of swapping is varied, the effect on
joint distributions of categorical variables is no longer understandable when
the data swapping procedure attempts to target at-risk individuals for swapping
using a simple targeting criterion.
\end{abstract}

\section{Introduction}

The U.S.\ Census Bureau is an important data source for government agencies and
researchers with questions regarding the population of the United States.
Providing these data with no safeguards against their misuse would be ideal for
researchers who want the most accurate data possible; however, the full
collected data cannot be released as this would endanger many residents. Data
confidentiality edits are employed by the Census Bureau in order to minimize the
possibility of potential adversaries using the data to identify and learn about
individuals in the data, while managing the tradeoff with statistical utility.
Data swapping is one such data confidentiality edit implemented by the U.S.\
Census Bureau. 

This paper attempts to study the effect that the Census Bureau's data swapping
algorithm has on the joint distribution of a pair of categorical variables. More
specifically, data swapping is simulated hundreds of times per swap rate for a
range of swap rates, and the Cram\'er's $V$ statistic is calculated for each
table generated from the swapped data. These tables are created from Census
Bureau data at a particular geography, e.g.\ the table of race by marital status
for a certain block in Pittsburgh, Pennsylvania. There are also two kinds of
data swapping performed in this analysis: non-targeted and targeted swapping,
with targeted swapping being a more complex procedure than non-targeted
swapping. 

Within academia, one obstacle in the way of a proper analysis of the effects
that data swapping has on statistical procedures is the necessary secrecy
surrounding it: the entire data swapping algorithm used by the Census Bureau is
not publicly accessible. However, several details of data swapping as it is
implemented at the U.S.\ Census Bureau are described by \citet{Griffin:1989fk,
Hawala:2003ty, Zayatz:2010yg}. Swapping is also in use at the Office of National
Statistics in the U.K.\; \citet{Shlomo:2011fk} investigate the effects of data
swapping by comparing their versions of non-targeted and targeted swapping.
Although the targeted data swapping scheme analyzed by Shlomo is unlikely to be
the swapping scheme in use at the U.S.\ Census Bureau, they share an important
property: ``at a higher geographical level and within control strata, the
marginal distributions are preserved.'' This property holds because data
swapping is done within a preset geographic level, i.e.\ within states, so
although the contingency tables generated at the block level are affected by
swapping, tables generated for sufficiently large areas (state and above) do not
need to be protected and are not changed at all by swapping. 

A case study done at the Census Bureau by \citet{Griffin:1989fk}, which
considered the full 1980 Census data on New Jersey, convinced the Bureau that
the distortion induced by data swapping is minimal. Since then, however, work by
\citet{Alexander21092010} has demonstrated a discrepancy in the distribution of
age by gender in the ACS Public Use Microdata Sample (PUMS) dataset, and data
swapping was determined to be a possible culprit. As they point out,
\begin{quote} Newer [disclosure avoidance techniques], such as swapping or
  blanking, retain detail and provide better protection of respondents'
  confidentiality. However, the effects of the new techniques are less
  transparent to data users and mistakes can easily be overlooked. Therefore
  these new techniques carry increased responsibility for both data users and
  data producers to vigilantly review the anonymized data. \end{quote} Most
recently, \citet{Crimi:2014qn} also noticed that data swapping may be affecting
the quality of estimates derived from the Census ACS PUMS dataset. 

One way this paper extends the Census Bureau's analyses is by studying the
effect of swapping on a statistical procedure that is already familiar to users
of the Census data: Pearson's chi-square statistic. Specifically, the Cram\'er's
$V$ values are compared between unswapped and swapped tables; Cram\'er's $V$ is
the the chi-square statistic but rescaled to be between 0 and 1. It is
calculated as $$V = \sqrt{\frac{\chi^2/n}{\min\{k-1, r-1\}}},$$ for a given
2-way $k$ by $r$ contingency table with total count $n$, where $\chi^2$ is the
chi-square statistic of the table. This particular measure was chosen because it
captures the joint distribution between two categorical variables, and it
simplifies the presentation of the results. The previously-cited work by
\citet{Shlomo:2011fk} also chose this measure of data utility, and demonsted
that their version of swapping decreased Cram\'er's $V$ values. Most recently, a
study by Census researchers \citet{lemonsdajaniyou2015} demonstrated the effect
of swapping on several common measures of statistical information, and at
varying swap rates, but it does not contrast targeted and non-targeted swapping
schemes. 

Other academic research into data swapping has involved analyzing variants of
the data swapping procedure, as this paper does. \citet{Lemons:2014ve} provides
a comprehensive account of the motivations and history of data swapping
research, and also contributes to the literature by analyzing the effect of a
variant of swapping done at three swap rates, from the perspective of
differential item functioning. \citet{Fienberg:2005vn} present a theoretical
justification for data swapping. \citet{Carlson2002} comprehensively analyze
rank-based swapping. For a more general overview of disclosure avoidance at the
Census Bureau, \citet{Zayatz:fk} provide a summary of benefits of different
data-masking procedures on microdata and tabular data. The targeted data
swapping algorithm presented in this paper is similar to that of the Census
Bureau in that it considers a set of variables to be preserved by swapping, and
another set to be emphasized for protection. The analysis will be done at a
comprehensive range of swap rates in order to capture the amount of instability
observed in the joint distributions of the swapped data.

\subsection{Data}
This analysis utilizes the PUMS and Summary Files, both derived from ACS data.
These are both public data products released by the U.S.\ Census Bureau. Table
\ref{tab:moresparse} is an example of a contingency table extracted from the ACS
Summary Files that contains at-risk cells. In this analysis, swapping is also
performed on ACS data that is only available on a need-to-know basis to Census
Bureau employees and external researchers, and can only be accessed at the
Census Bureau or at Research Data Centers. These data are desirable since they
contain full geographic information for all records, and therefore the correct
joint distributions between the variables. There is a possibility of the data
swapping procedure applied to the ACS data differing from what was applied to
the Decennial Census data, but in a publication describing the application of
disclosure avoidance to Census 2010 and American Community Survey data,
\citet{Zayatz:2010yg} claim that the ``procedures will remain virtually the same
as those used for Census 2000 sample long form data.'' 

\begin{table}
  \centering
  \begin{tabular}{r|rrrrr}
    \hline
    Age & Married & Widowed & Divorced & Separated & Never Married \\ 
    \hline
    $\leq 16$ &   3 &   0 &   0 &   0 &  30 \\ 
    17 &   0 &   0 &   0 &   0 &  35 \\ 
    18 &   $1^*$ &   0 &   0 &   0 & 194 \\ 
    19 &   $1^*$  &   $1^*$ &   0 &   0 & 269 \\ 
    20 &   $1^*$ &   0 &   0 &   0 & 188 \\ 
    21 &   4 &   0 &   0 &   $1^*$ & 153 \\ 
    22 &   $2^*$ &   0 &   0 &   0 & 111 \\ 
    23 &   $2^*$ &   0 &   0 &   0 &  80 \\ 
    24 &  14 &   0 &   $1^*$ &   $2^*$ &  58 \\ 
    $\vdots$ & $\vdots$ & $\vdots$ & $\vdots$ & $\vdots$ & $\vdots$ \\
    93 &   0 &   9 &   0 &   0 &   $2^*$ \\ 
    $\geq 94$ &   3 &  16 &   $1^*$ &   0 &   $1^*$ \\ 
    \hline
  \end{tabular}
  \caption{A table extracted from the Census Bureau ACS Summary Files which
  contains a significant amount of at-risk cells, denoted by $1^*$ and $2^*$.}
  \label{tab:moresparse}
\end{table}

\subsubsection{Dummy data}
Simplified simulations are first run on a dummy dataset that is not generated
from any Census Bureau data, so as to establish some expectations for the more
complex simulations on real data. The dummy dataset has as variables Age,
Income, and Tract, and is designed to mimic a dataset with tract-level
geography. For each tract $t$, a parameter $b_t$ is randomly generated so that
each observation is generated by $\mathrm{Income} = b_t \mathrm{Age} +
\epsilon$, where $\epsilon$ are independent noise terms from a Poisson
distribution, so as to create a geography-dependent correlation structure in the
dummy data. Individuals were assigned to 50 of these artificial Tracts, and
there are 200 individuals in each Tract. They are all assumed to be in the same
PUMA, so that swapping can happen freely between the Tracts. 

In the dummy data, 62\% of Tracts have a Cram\'er's $V$ value generated from the
table of Poor vs.\ Young lower than that of the ``combined table'' where we
marginalize over Tract (Poor is just a binary variable defined so that about
23.5\% of individuals in the dummy data are considered "poor", and similarity
for Young, so that about 32.7\% are "young"). In other words, the data are such
that most of the Tracts corresponded to Cram\'er's $V$ values that are lower
than that of the total contingency table generated by ignoring Tract
information.

\subsubsection{ACS data}

The Census Bureau's American Community Survey is designed to collect data beyond
the few variables that are collected in the decennial census. The analysis in
this paper is based on the ACS data that was collected between 2007--2011. The
5-year ACS data are used, as opposed to the 1-year or 3-year data, because the
5-year data includes finer geography and the greatest number of individual
records. About 230 variables are provided in the public-use version of these
data. 

Publicly-available Summary Files from the ACS include contingency tables for
certain combinations of variables. These public datasets are all created from
data that have been swapped prior to the generation of any Summary Files and
PUMS. There are potentially billions of different contingency tables available
(for thousands of combinations of variables at many geography levels for all
regions of the U.S.). The data in this study are restricted to be from Allegheny
County, PA, which is a large enough county that it contains its own Public Use
Microdata Areas (PUMAs, which are themselves always larger than census tracts).
For this analysis, it will be assumed that any pairwise swap is performed
between two households in the same PUMA. 

The full, unswapped Census data can only be accessed by certain authorized
researchers, contractors, and employees of the Bureau. Since the dataset must
remain on Census servers, it cannot be taken out of the Bureau's headquarters or
the handful of Research Development Centers (RDCs). This dataset is useful since
it contains full geographic information and full information within the survey
variables. Specifically, for this analysis, knowledge of the actual tract that
each individual resides in is necessary (this level of geographic detail is not
provided in the PUMS dataset). Knowing the actual tracts means that the data
will have accurate joint and geographic structure. Using just the PUMS, one can
only generate synthetic data that will not contain correct joint distributions.

\section{Methodology}

\subsection{Non-targeted versus targeted swapping}
For this paper, non-targeted swapping will refer to swapping in the simplest
case; the algorithm for this is given in Algorithm \ref{alg:randswap}. Note that
any swapped pair is required to to match on some predetermined set of
attributes, but any household is as likely as any other to be swapped. No
attempt is made to target the individuals who are likely to appear as a one in a
contingency table generated from these data. 

\begin{algorithm}
\KwData{Data frame of $n$ individual records, $s$ swap rate}
\KwResult{Swapped data}
randomly select $n \times s$ individual records\;
randomize the order of those records\;
\For{$i$ in 1:$n$}{
  \If{Record $i$ is unswapped}{
    Give record $i$ the physical address of another unswapped record in a
    household with the same number of people, given that they match on a set
    of attributes $S$\;
    Give the other unswapped record the address of record $i$\;
    Do the same for the rest of the household members\;
  }
}
\caption{Pseudocode for the non-targeted swapping algorithm. }
\label{alg:randswap}
\end{algorithm}

\begin{algorithm}
\KwData{Data frame of $n$ individual records, $s$ swap rate}
\KwResult{Swapped data}
calculate the disclosure risk score for each record\;
select only the top $n \times s$ records in terms of
disclosure risk score\;
randomize the order of those records\;
\For{$i$ in 1:$n$}{
  \If{Record $i$ is unswapped}{
    Give record $i$ the physical address of another unswapped record in a
    household with the same number of people, given that they match on a set
    of attributes $S$\;
    Give the other unswapped record the address of record $i$\;
    Do the same for the rest of the household members\;
  }
}
\caption{Pseudocode for the targeted swapping algorithm, which is similar to
  the non-targeted swapping algorithm, except that the subset of individuals
chosen for swapping have been chosen due to their disclosure risk score. }
\label{alg:targswap}
\end{algorithm}

Psuedocode for targeted swapping is given in Algorithm \ref{alg:targswap}.
Intuitively, individuals who are most at risk for identification should be the
focus of swapping, e.g.\ especially rich or old individuals. They can be
protected by swapping them away from their current geographic areas since they
are more likely to be ones in a contingency table for a suitably small area. The
procedure for the targeting scheme analyzed in this study is to count the number
of variables for which an individual has an extreme value. For instance, if an
individual is in the top $q$th quantile of income and in the bottom $q$th
quantile for number of toilets owned, but all other variables are outside of
their respective top and bottom $q$th quantiles, then their count is two, since
only two of their attributes is considered to be at-risk by this measure. 

\begin{figure}
\centering
\begin{adjustbox}{addcode={\begin{minipage}{\width}}{\caption{
        The two plots show the mean proportion (and standard error bars) of ones
        being changed to different values by swapping in the contingency table
        of age by marital status for one particular tract. Non-targeted (left)
        and targeted (right) swapping are shown. 200 simulations of each
        swapping scheme were done for each swap rate depicted to generate the
        above plots. 
  }
\label{fig:logcompare} 
\end{minipage}},rotate=90,center}
\includegraphics[width=0.6\textwidth]{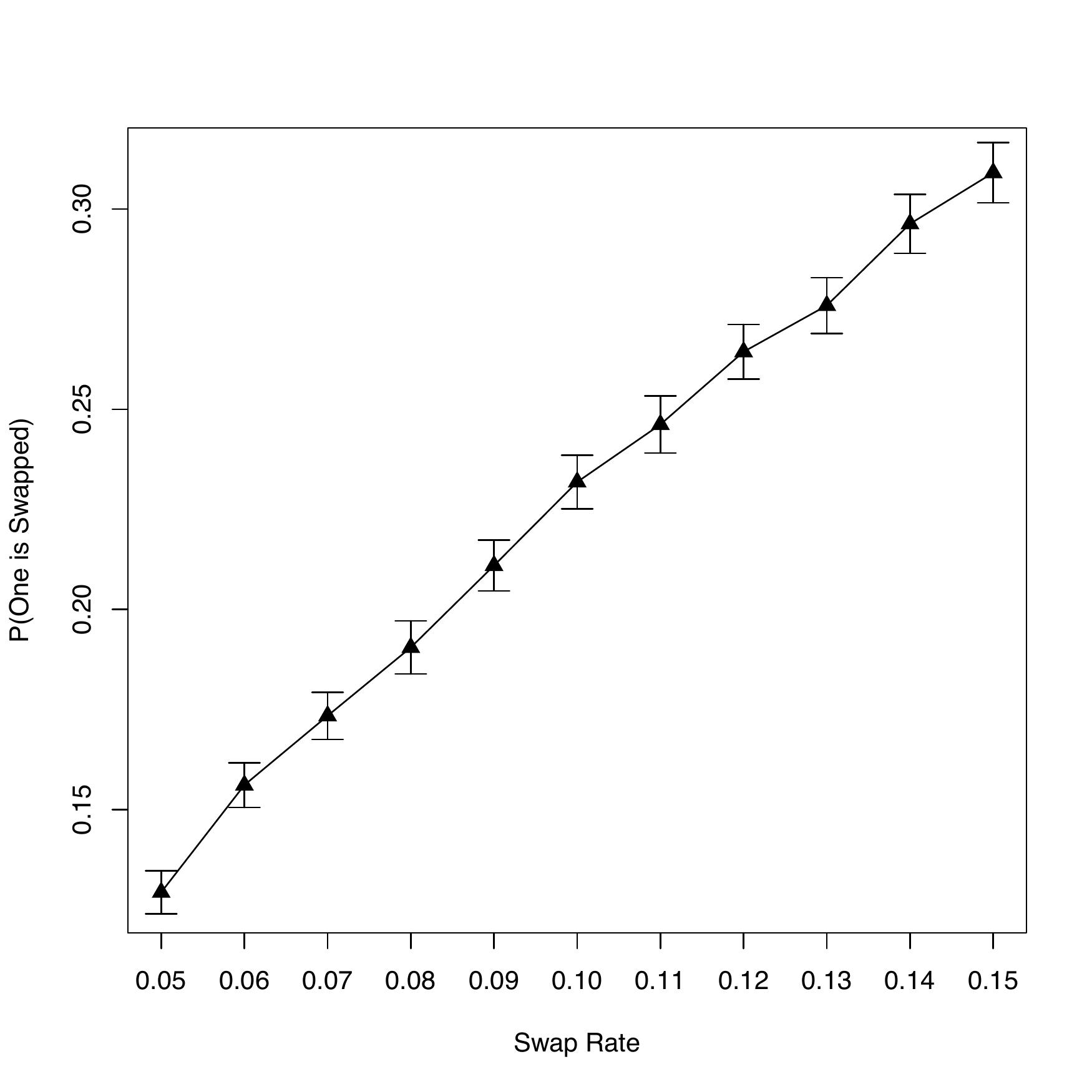}
\includegraphics[width=0.6\textwidth]{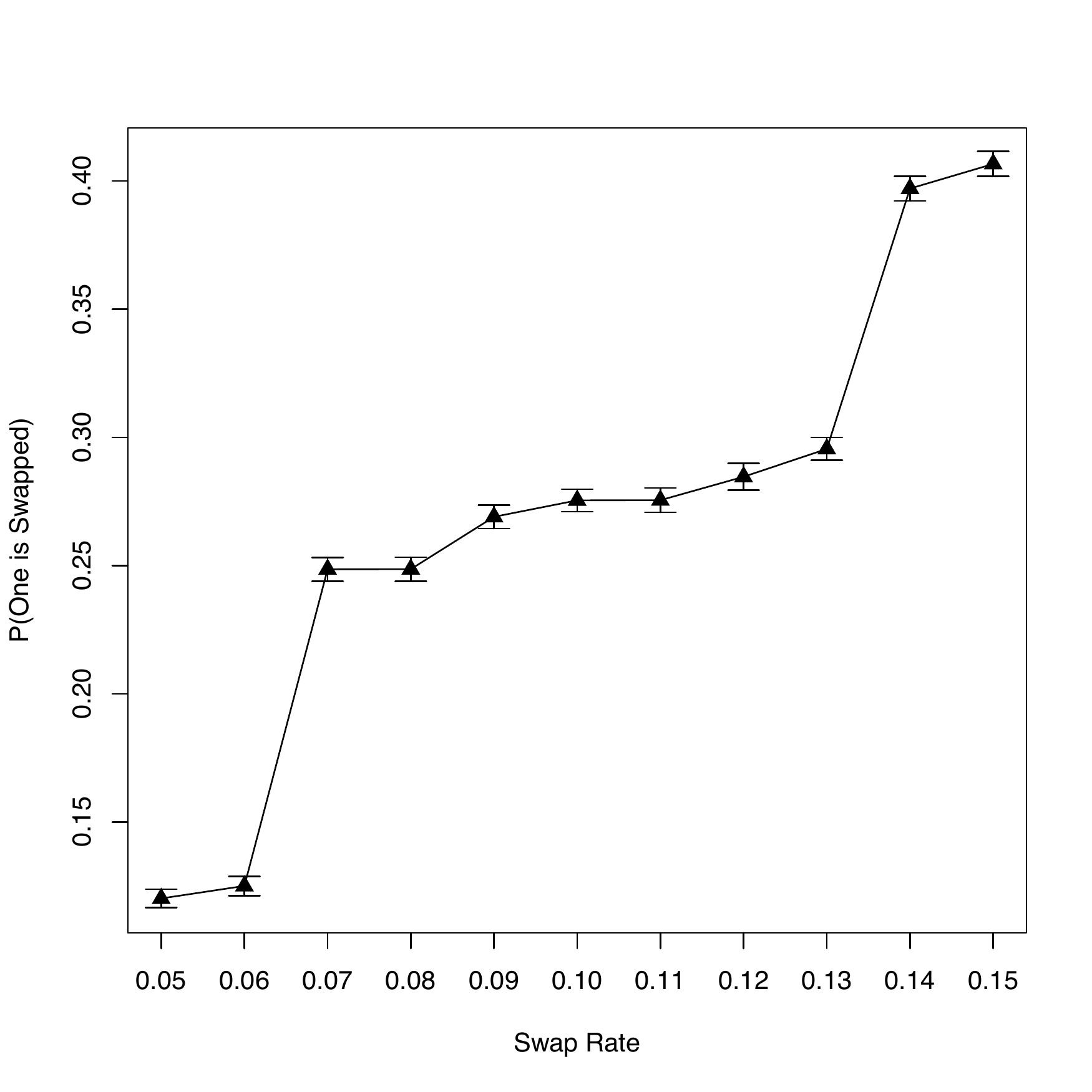}
\end{adjustbox}
\end{figure}

The targeting scheme used for this analysis ranks individuals according to the
log of the relative frequencies of their variable values. For example, if in a
dataset only 10 out of 50 individuals are married, and 25 out of 50 individuals
are male, then a married male's ``disclosure risk score'' is $\log(0.2) +
\log(0.5)$. The lower the score, the more likely one is to be at risk for
disclosure. When $m$ people are chosen for swapping, these are really the $m$
individuals with the $m$ lowest scores. Figure \ref{fig:logcompare} demonstrates
the effectiveness of the targeting procedure in protecting against potential
unique combinations of levels, i.e.\ ones in a contingency table. These
simulations were performed on the 5-year ACS PUMS with synthetically-generated
tract information. 

\subsection{Analyzing the effect of data swapping with simulations}
By simulating the data swapping procedure hundreds of times, effect of swapping
on Cram\'er's $V$ can be estimated as the swap rate is varied. The randomness in
the simulations comes from the uncertainty in which compatible match is made for
each pairwise swap. These simulated values are shown as plots of the average
Cram\'er's $V$ by swap rate. In addition, standard error bars are added, based
on the sample standard deviation and the number of simulations; QQ plots of the
observed Cram\'er's $V$ values at each swap rate strongly indicated normality,
justifying the use of normal error bars. 

Within the simulations, swapping is done at a range of swap rates, typically
between 5\% and 15\%. The granularity of the range of simulated swap rates
varies as the simulations sometimes take on the order of days to run. Therefore,
when generating plots that do not require very fine increments of the swap rate,
I choose a small number of swap rates between 5\% and 15\%.

\section{Results}
\subsection{Results from simplified simulations on dummy data}

\begin{figure}
\centering
\begin{adjustbox}{addcode={\begin{minipage}{\width}}{\caption{
For the leftmost plot, each line represents one tract's change in Cram\'er's $V$
value for increasing swap rates. Each value plotted here represents the mean
over 150 simulations of the data swapping procedure. The red paths represent
tracts which had a lower Cram\'er's $V$ value than the entire PUMA; likewise,
the blue paths represent tracts with a higher value. 
For the rightmost plot, each trajectory represents one tract's Cram\'er's $V$
values as the swap rate is increased from 0 to 1 (i.e., from 0\% to 100\%). The
green dashed line represents the mean Cram\'er's $V$ value across all of the
unswapped tracts. The convergence of the trajectories is evidently towards the
green line. Both plots were generated with the same data, but only 50
simulations were done for each combination of tract and swap rate on the
rightmost plot. 
  }
\label{fig:inducedDiff} 
\end{minipage}},rotate=90,center}
\includegraphics[scale=0.5]{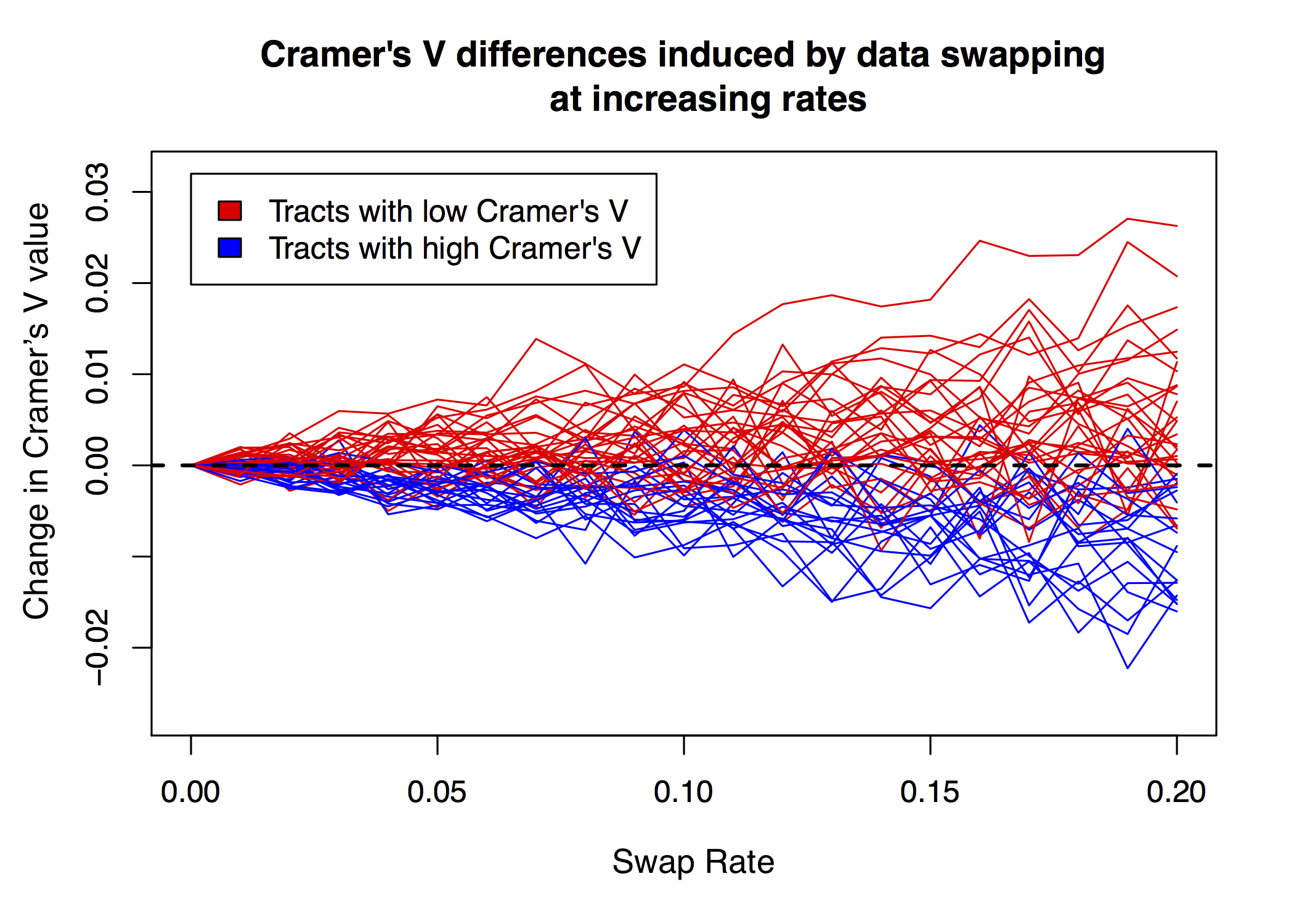}
\includegraphics[scale=0.5]{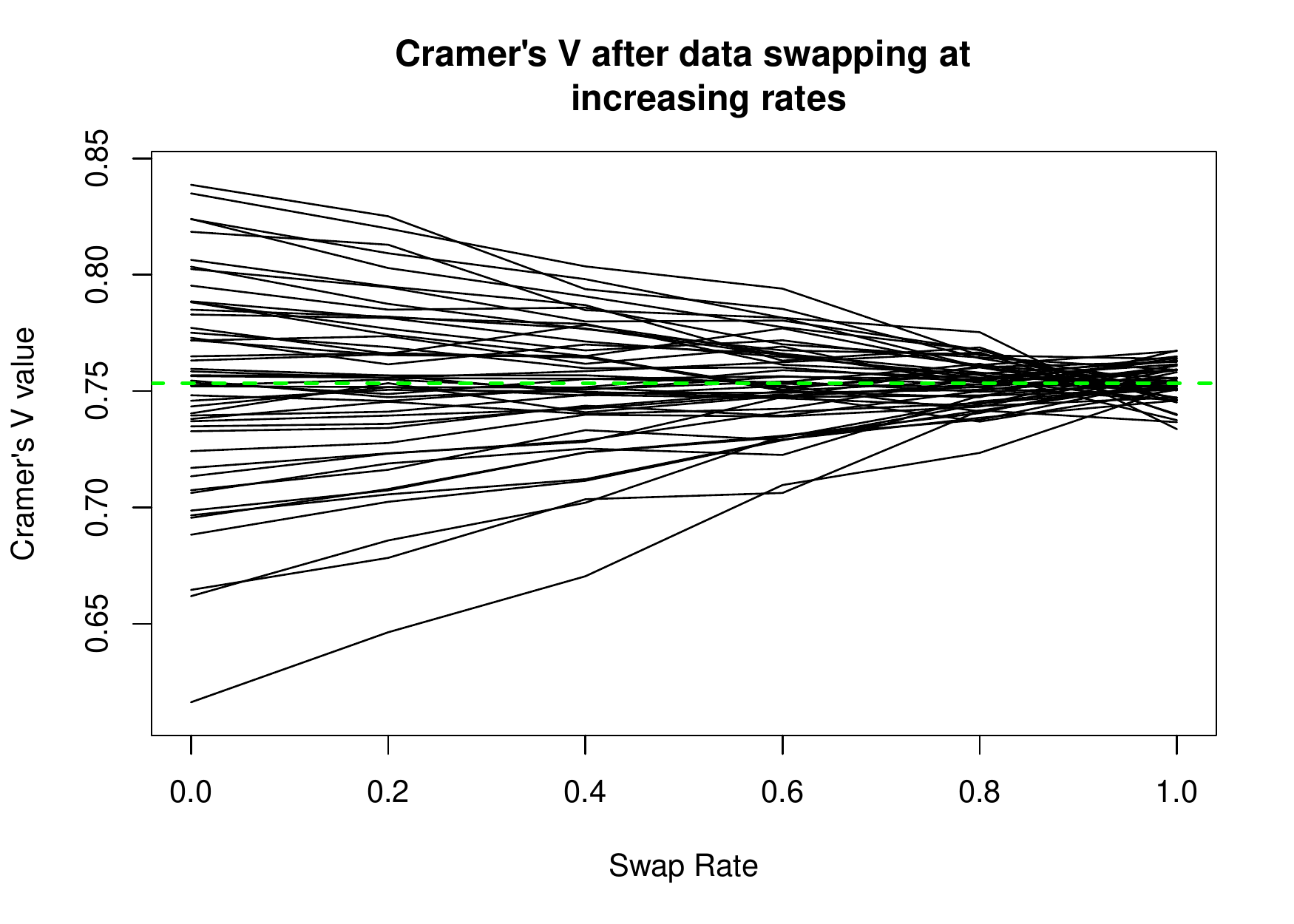}
\end{adjustbox}
\end{figure}

Using the dummy data, 150 simulations of a simplified swapping procedure were
performed for 21 different swap rates (0\%, 1\%, 2\%, \dots, 20\%). This version
of swapping does not attempt to match pairs of households or target at-risk
households. In Figure \ref{fig:inducedDiff}, as the swap rate is increased from
0\% (no swapping) to 20\%, the tracts that originally had a low (high)
Cram\'er's $V$ value  ended up with higher (lower) Cram\'er's $V$ values as the
swap rate increased. Low and high Cram\'er's $V$ values correspond to values
lower and higher than the Cram\'er's $V$ value of the contingency table of the
entire PUMA, respectively. Every tract has its Cram\'er's $V$ value increasingly
nudged towards a central value as the swap rate is increased: the clear
separation in the red and blue trajectories demonstrates this. Figure
\ref{fig:inducedDiff} also demonstrates that these trajectories of Cram\'er's
$V$ values appear to be tending towards the average Cram\'er's $V$ value across
all of the PUMA's tracts. 

\subsection{Results from simulations on the synthetic data}
Recall that the synthetic dataset was generated from U.S.\ Census Bureau data,
but with artificial tract-level geography since only PUMA-level information is
provided in the public dataset. Two data swapping algorithms, the non-targeted
and the targeted swaps, were applied to this dataset. In both cases, some
variables were controlled for by ensuring that any two swapped households had to
match on these variables. This differs from the more simplistic non-matched
swapping that was performed while swapping the dummy data; note that a
comparable matching stage is in the Census Bureau's own version of swapping. 

\begin{figure}
\centering
\begin{adjustbox}{addcode={\begin{minipage}{\width}}{\caption{
The plots of Cram\'er's $V$ by swap rate for two different artificial tracts.
The curves are tending towards a central Cram\'er's $V$ value. These two plots
were generated on the synthetic dataset, and non-targeted swapping was used. The
plots for the rest of the tracts essentially look the same; the story here is
that some of the tracts experience a lowering of Cram\'er's $V$ values, and some
tracts experience the opposite, as the swap rate is increased. }
\label{fig:unexpectedrand} 
\end{minipage}},rotate=90,center}
\graphicspath{{sim_images/random/}}
\includegraphics[scale=0.48]{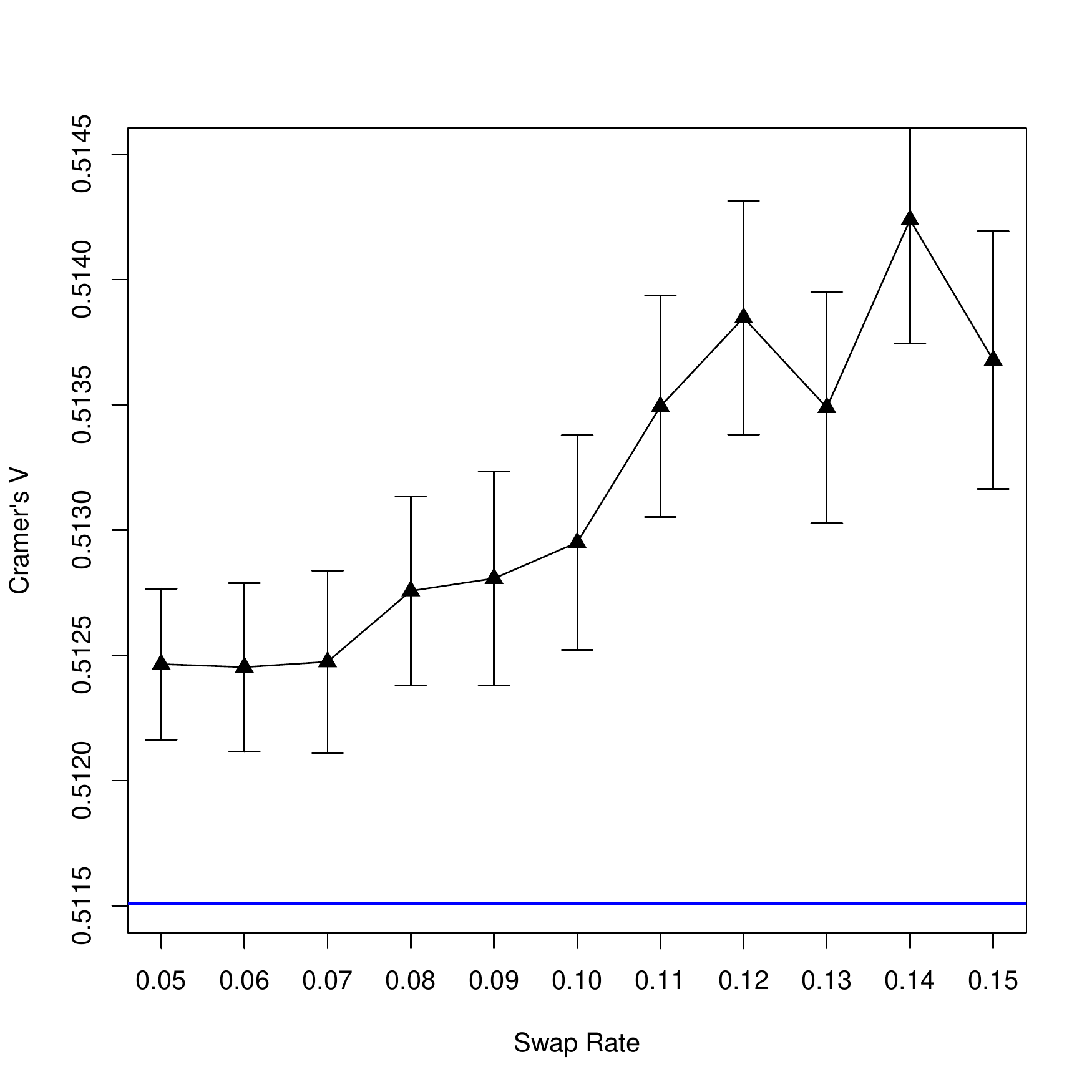}
\includegraphics[scale=0.48]{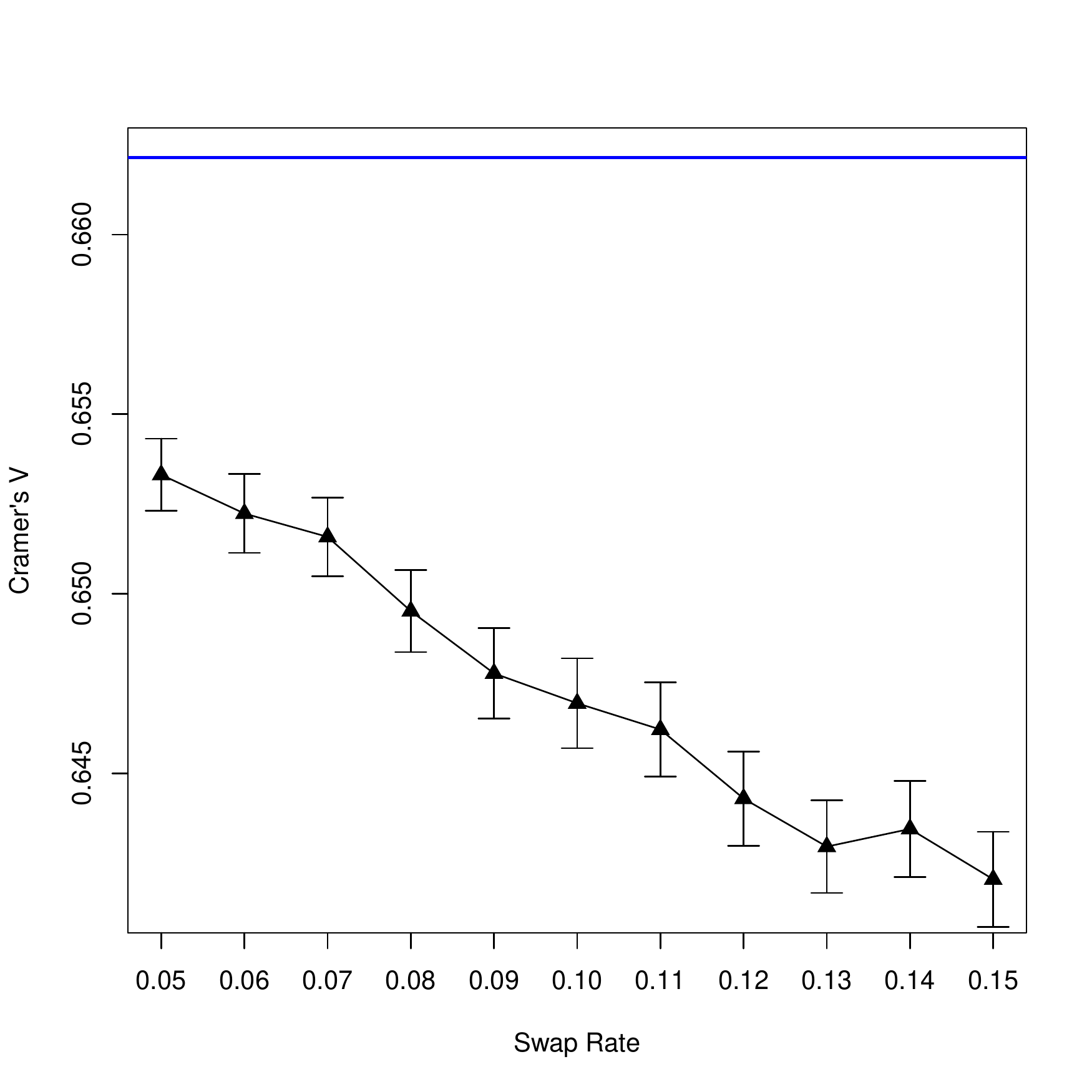}
\end{adjustbox}
\end{figure}

The trajectory of the Cram\'er's $V$ values for the table of age and marital
status as the swap is varied is plotted for two representative tracts in Figure
\ref{fig:unexpectedrand}. These are plots of the average Cram\'er's $V$ versus
swap rate. For instance, for the left plot, the trajectory represents the
average Cram\'er's $V$ (plus standard error bars) after performing non-targeted
swapping 1,000 times at each swap rate, for the table of age and marital status
for a particular tract. To explain the effect of the simplest case of data
swapping, the non-targeted swap, note that the Cram\'er's $V$ values of the
different tracts seem to be converging to a common point as the swap rate
increases. The blue horizontal lines denote the Cram\'er's $V$ values for the
unswapped tables, and the trajectories are moving away. Evidently, they are
converging towards something like the average Cram\'er's $V$ value across all of
the tracts, as was observed in the analysis of the dummy data. In other words,
the inclusion of a matching stage is not noticeably changing the nature of data
swapping's effect on joint distributions. 

\begin{figure}
\centering
\begin{adjustbox}{addcode={\begin{minipage}{\width}}{\caption{
Same setup as Figure \ref{fig:unexpectedrand}, with the same two tracts, but
with targeted swapping. In this case, there does not appear to be any general or
explainable trend in the trajectories. }
\label{fig:unexpectedtarg} 
\end{minipage}},rotate=90,center}
\graphicspath{{sim_images/target/}}
\includegraphics[scale=0.48]{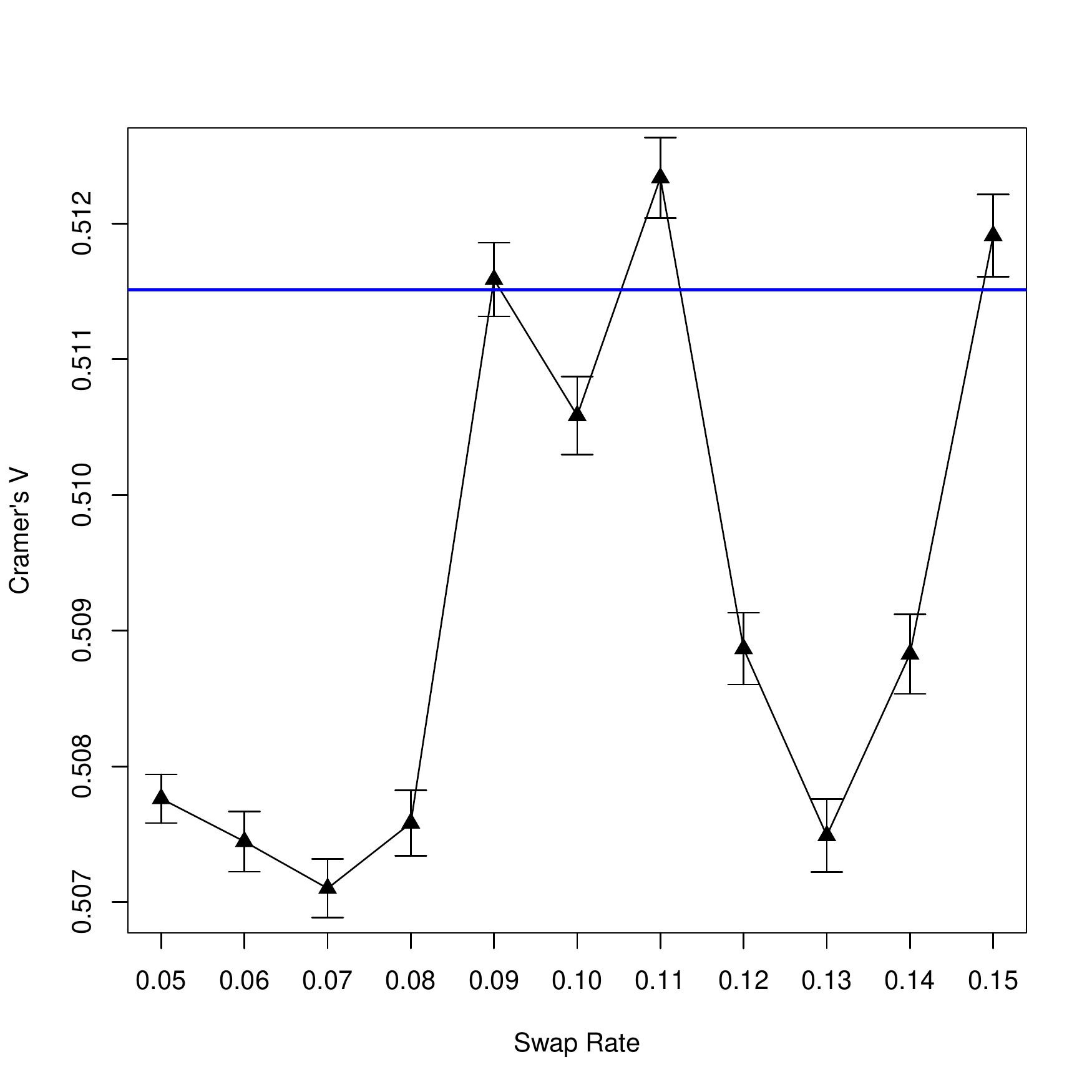}
\includegraphics[scale=0.48]{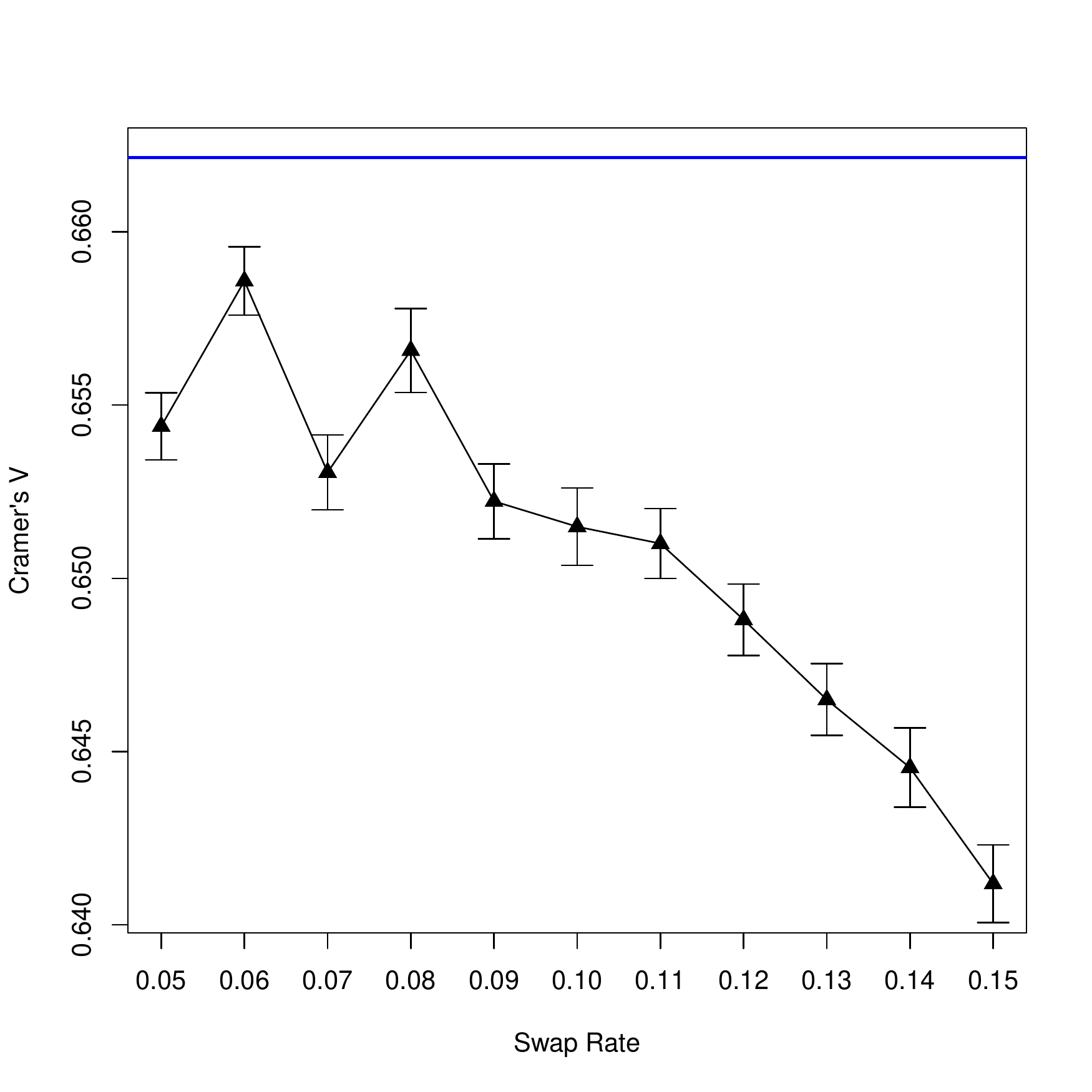}
\end{adjustbox}
\end{figure}

\begin{figure}
\centering
\begin{adjustbox}{addcode={\begin{minipage}{\width}}{\caption{
Same setup as Figure \ref{fig:unexpectedtarg}, with the same two tracts, except
that the age variable is now binned so that it only has two levels, so as to
minimize the potential for instability in the Cram\'er's $V$ values due to small
table counts. While the plots are no longer the same as Figure
\ref{fig:unexpectedtarg}, there is still no explainable trend here. }
\label{fig:unexpectedbin} 
\end{minipage}},rotate=90,center}
\graphicspath{{sim_images/targetbin/}}
\includegraphics[scale=0.48]{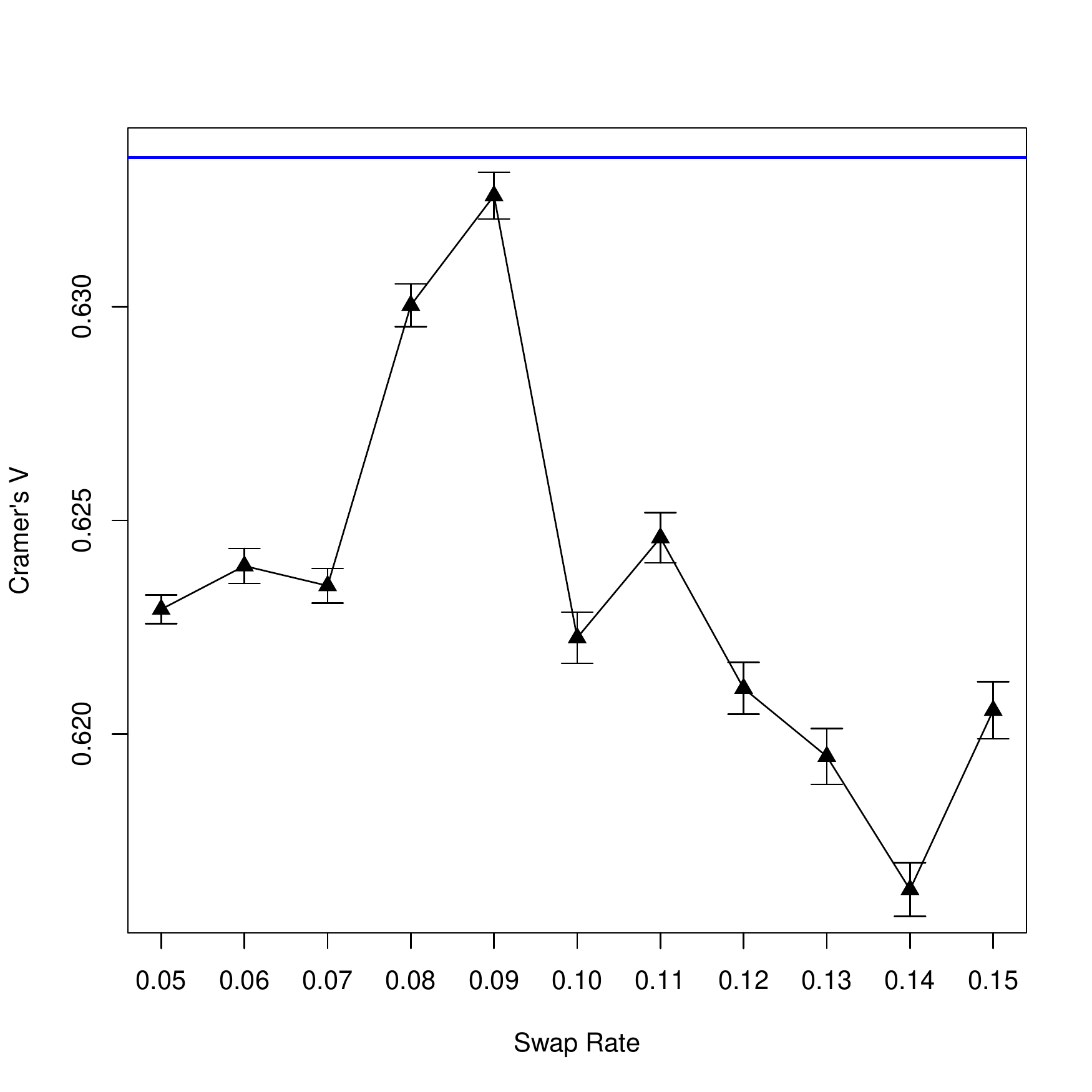}
\includegraphics[scale=0.48]{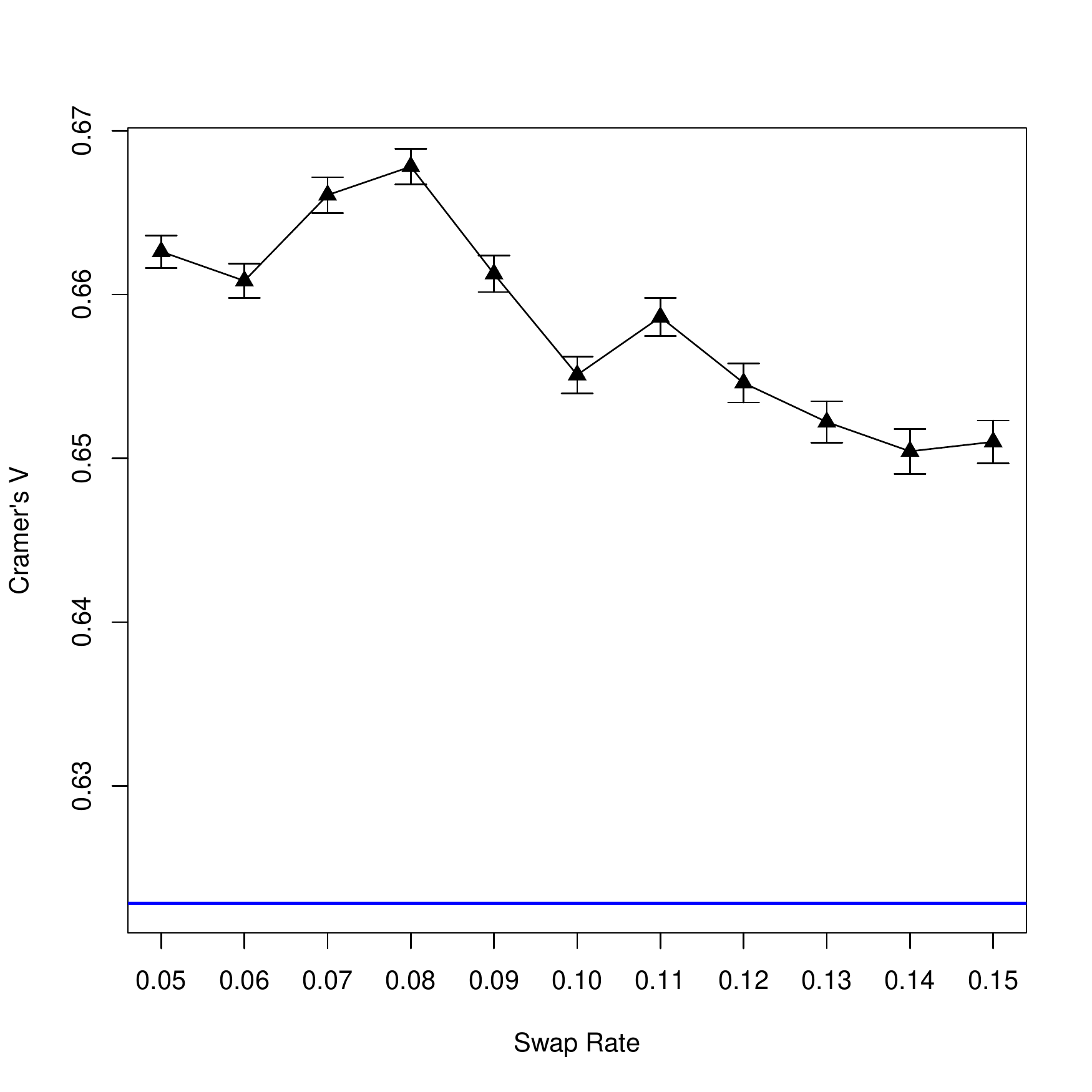}
\end{adjustbox}
\end{figure}

The same was done for the targeted swapping procedure, as shown in Figure
\ref{fig:unexpectedtarg}. However, it no longer seems that the trajectories are
tending towards a central value, at least not at this range of swap rates. Both
procedures were identical except for the additional targeting stage. In both
cases, gender and marital status were designated as matching variables, so that
no two individuals with non-matching marital statuses and genders would be
selected as swap candidates. By adding a simple and generic stage for targeting
the individuals most at-risk for swapping, the joint distributions output by the
swapping algorithm can no longer be reliably characterized or predicted.
Furthermore, this effect does not seem to be attributable to instability in
Cram\'er's $V$ induced by sparsity in the tables since there is still as little
explainable trend if the values are calculated from a condensed table where age
is binned into only two levels so that the table contains only ten entries, as
is seen in Figure \ref{fig:unexpectedbin}.

\subsection{Results from simulations on the unswapped data}
Unlike the publicly available data, the unswapped data at the Bureau contain the
true joint distributions within the collected data. Since the goal of this
analysis was to study the effect of swapping on true underlying associations
between variables, it was important to perform swapping simulations on this
pristine dataset. 

\begin{figure}
\centering
\includegraphics[width=0.49\textwidth]{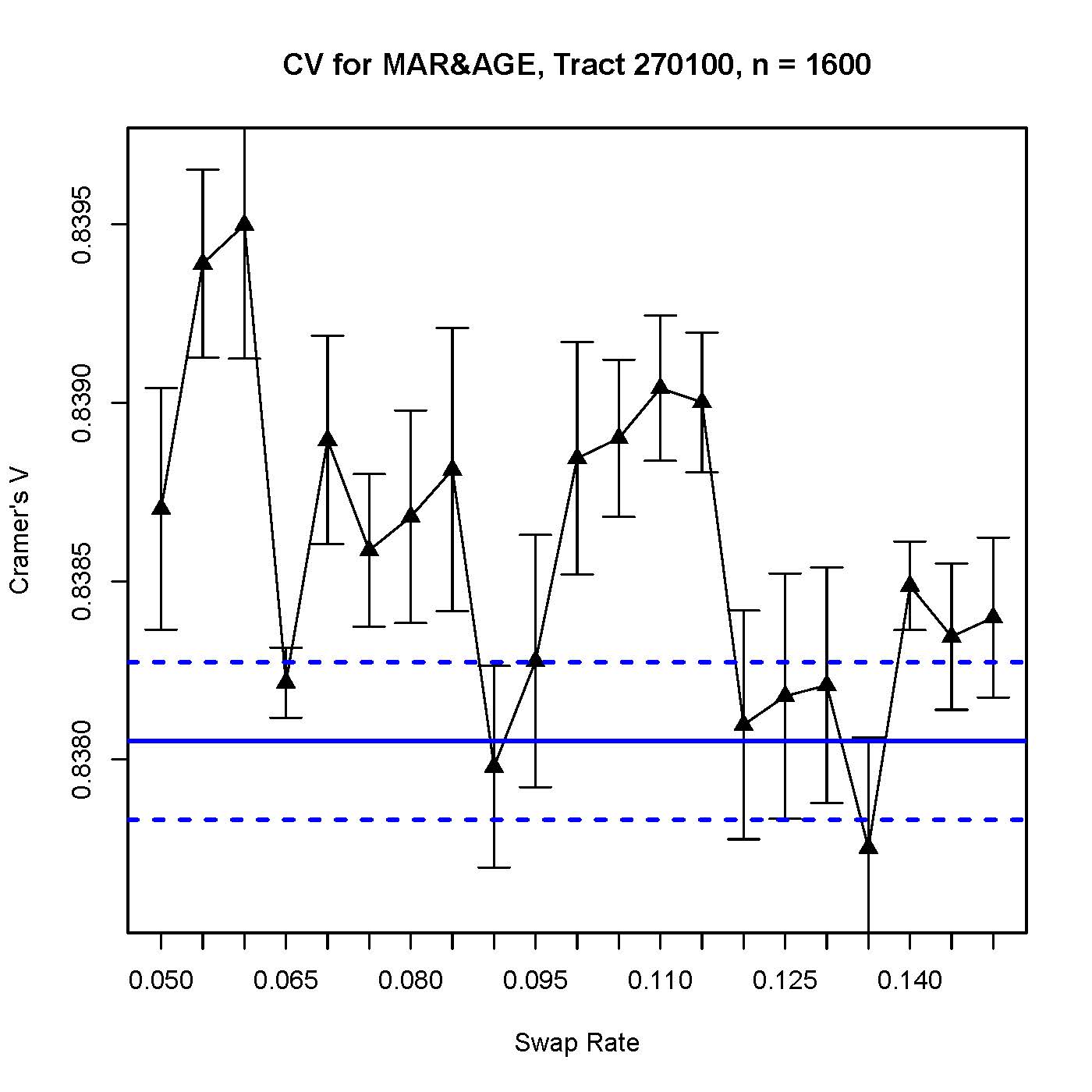}
\includegraphics[width=0.49\textwidth]{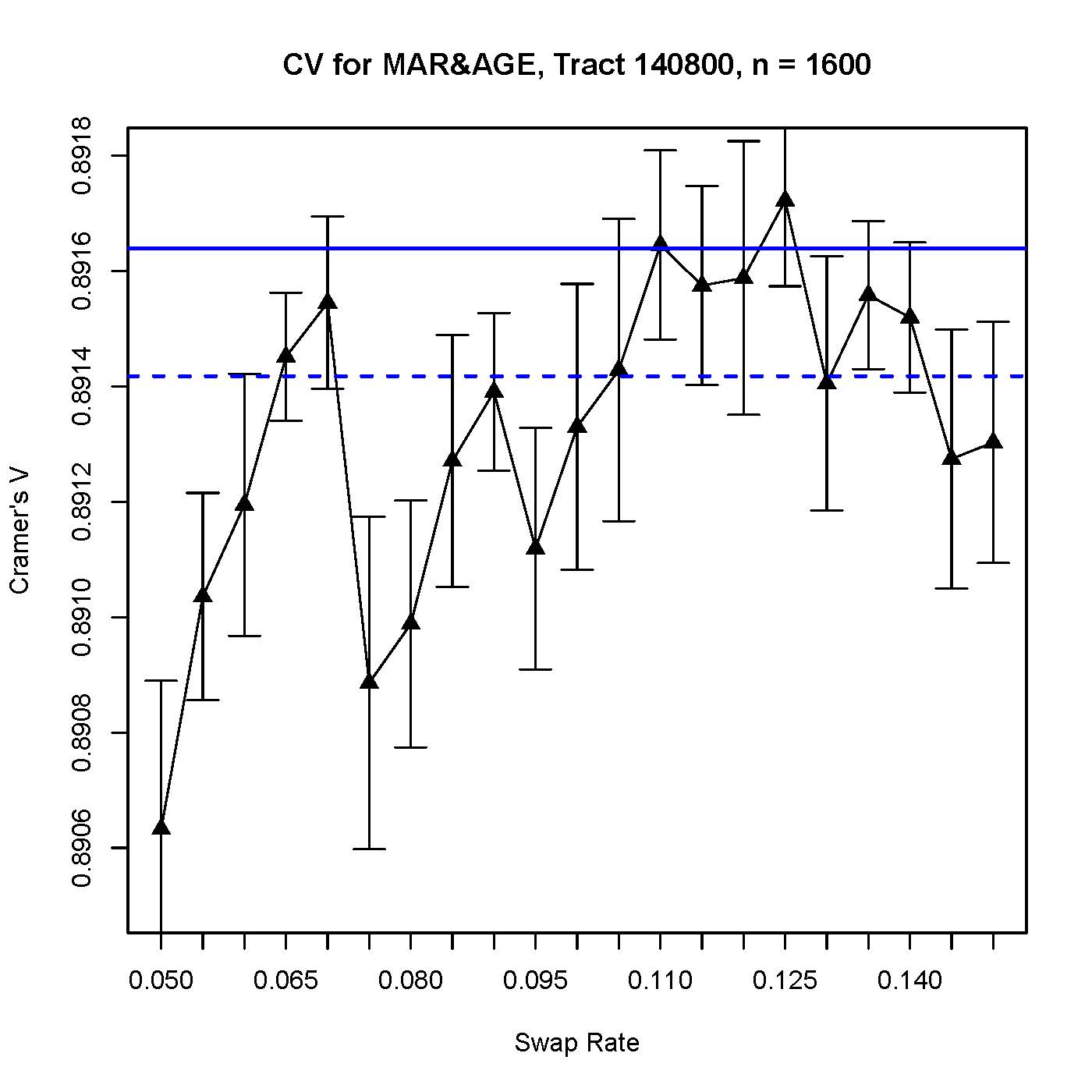} \\
\includegraphics[width=0.49\textwidth]{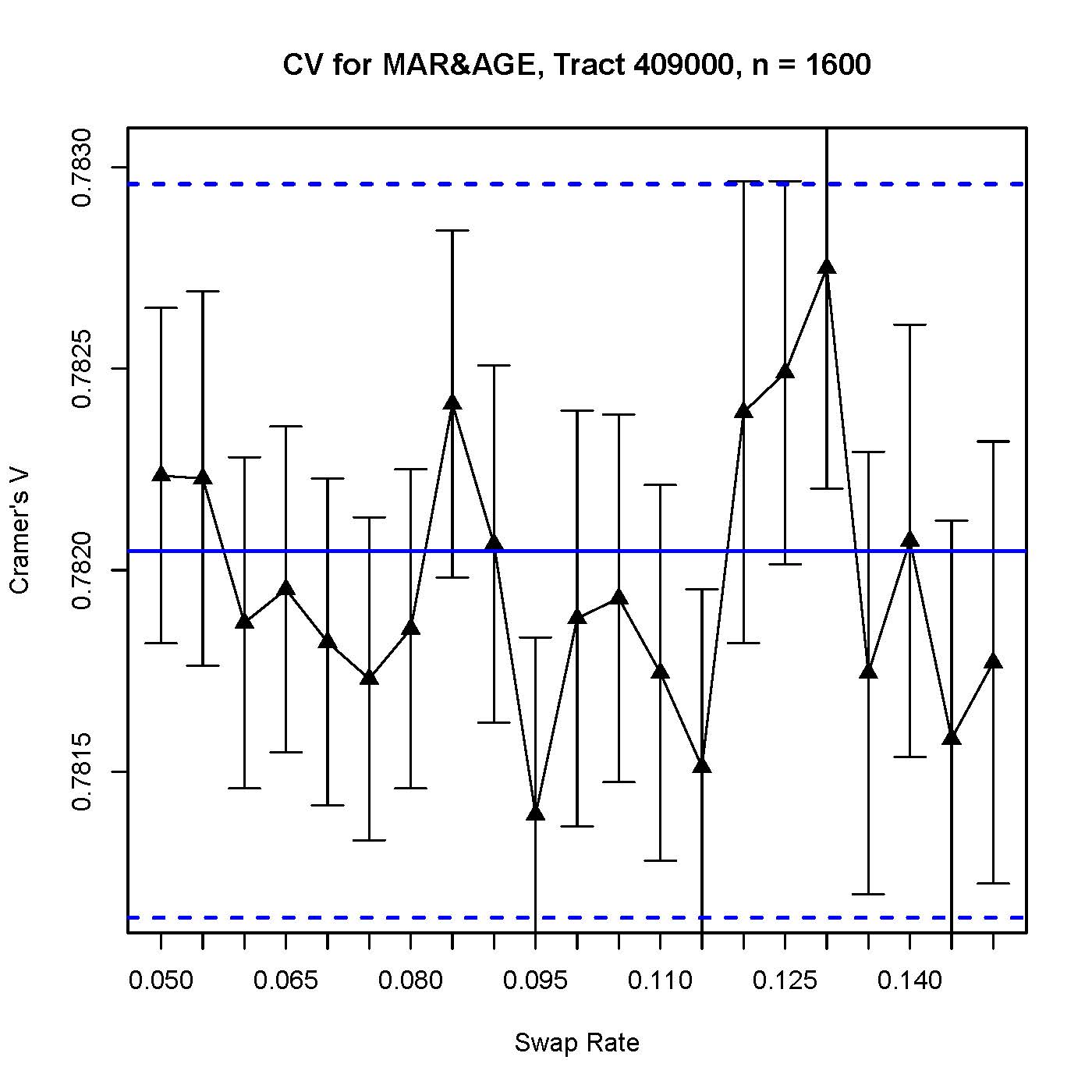}
\includegraphics[width=0.49\textwidth]{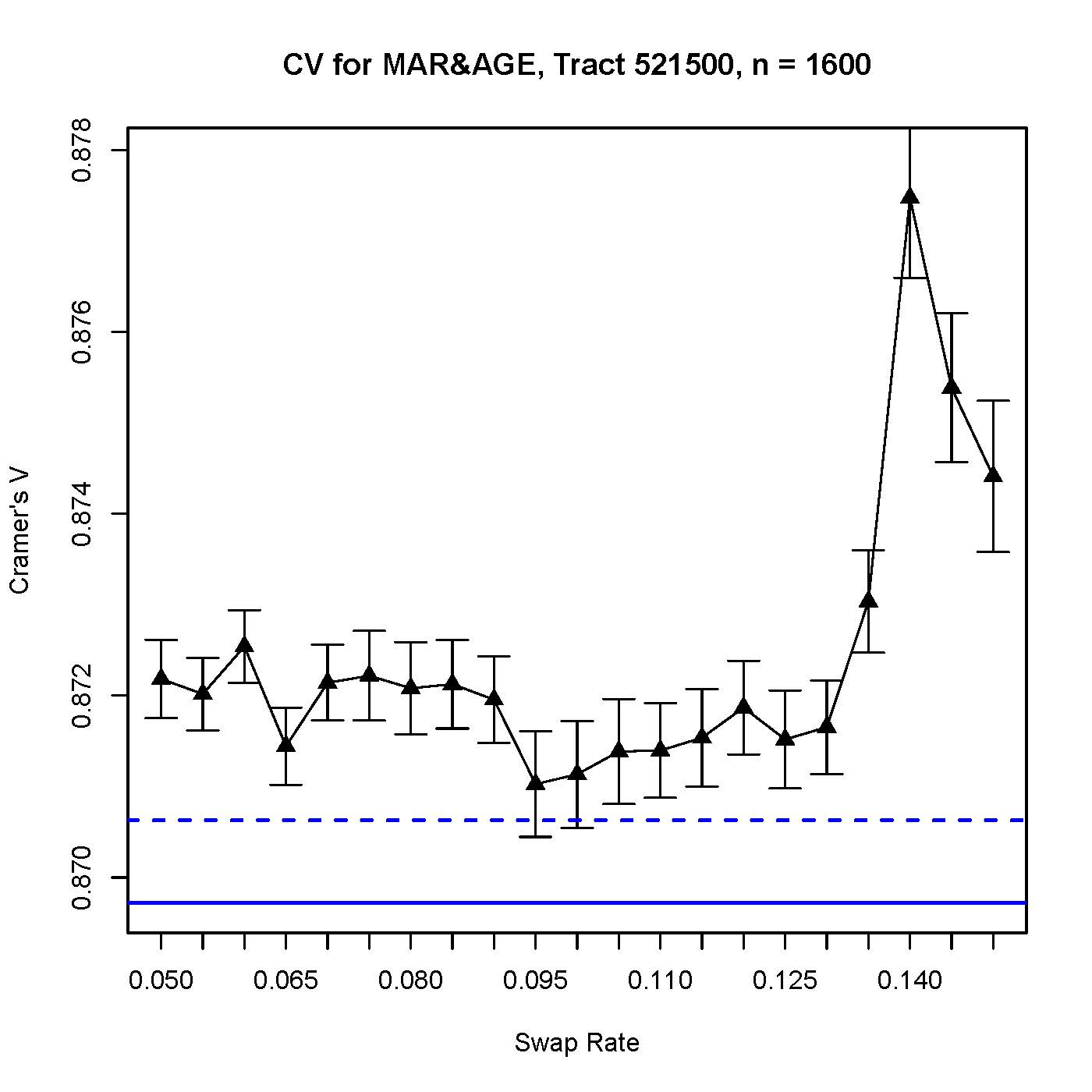}
\caption{ Cram\'er's $V$ versus swap rates for the tables of marital status by
age (very sparse but strong association). }
\label{fig:unswapplots} 
\end{figure}

\begin{figure}
\centering
\includegraphics[width=0.49\textwidth]{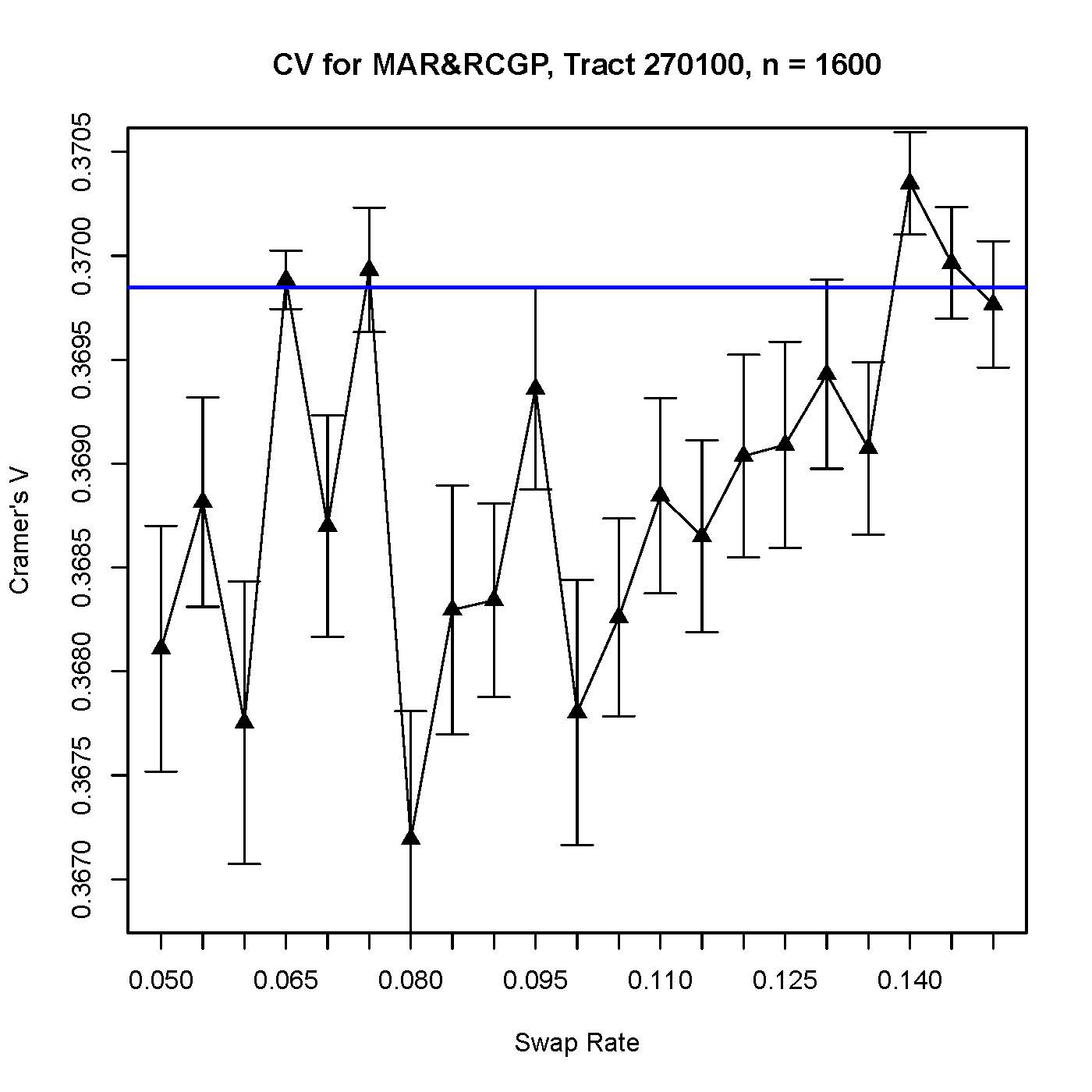}
\includegraphics[width=0.49\textwidth]{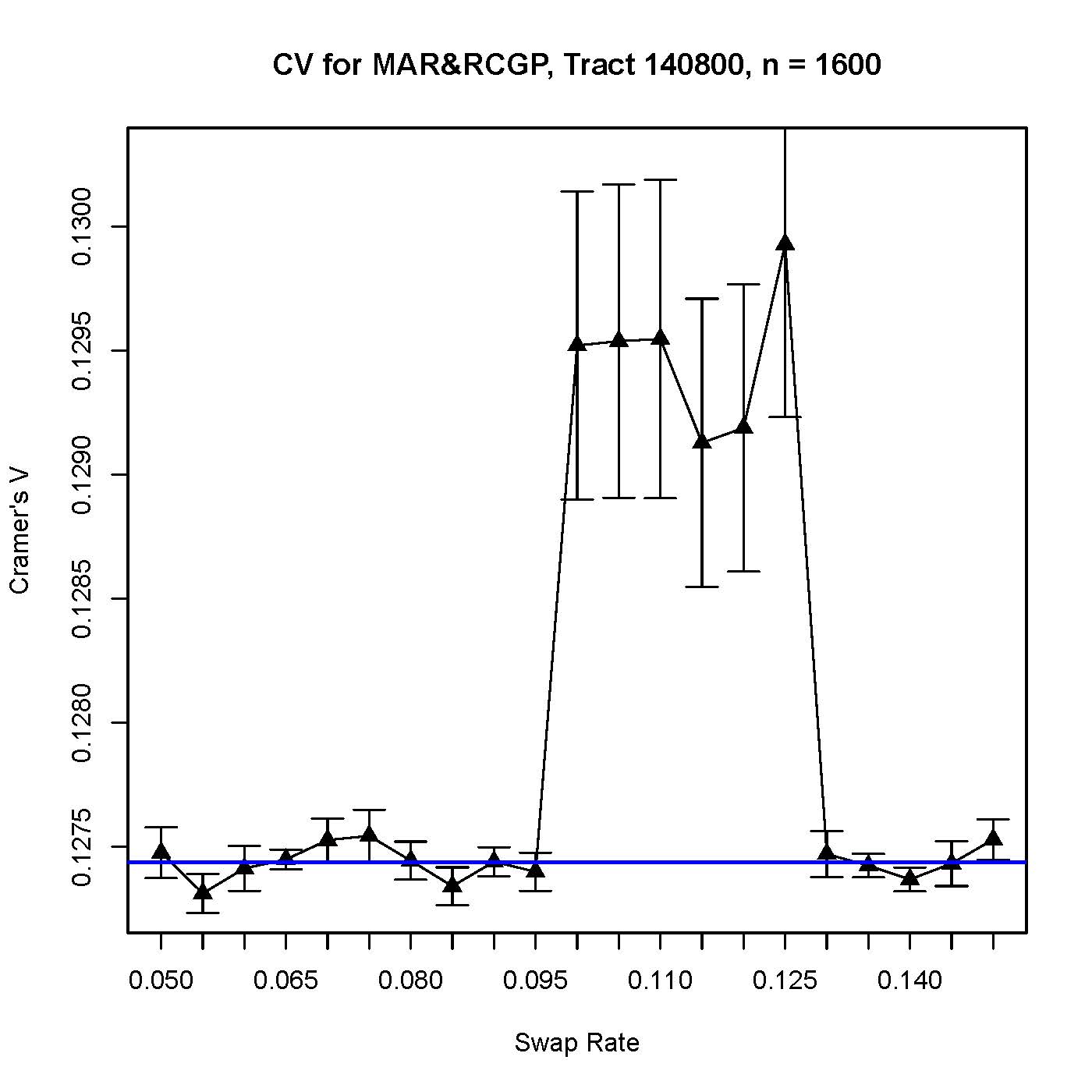} \\
\includegraphics[width=0.49\textwidth]{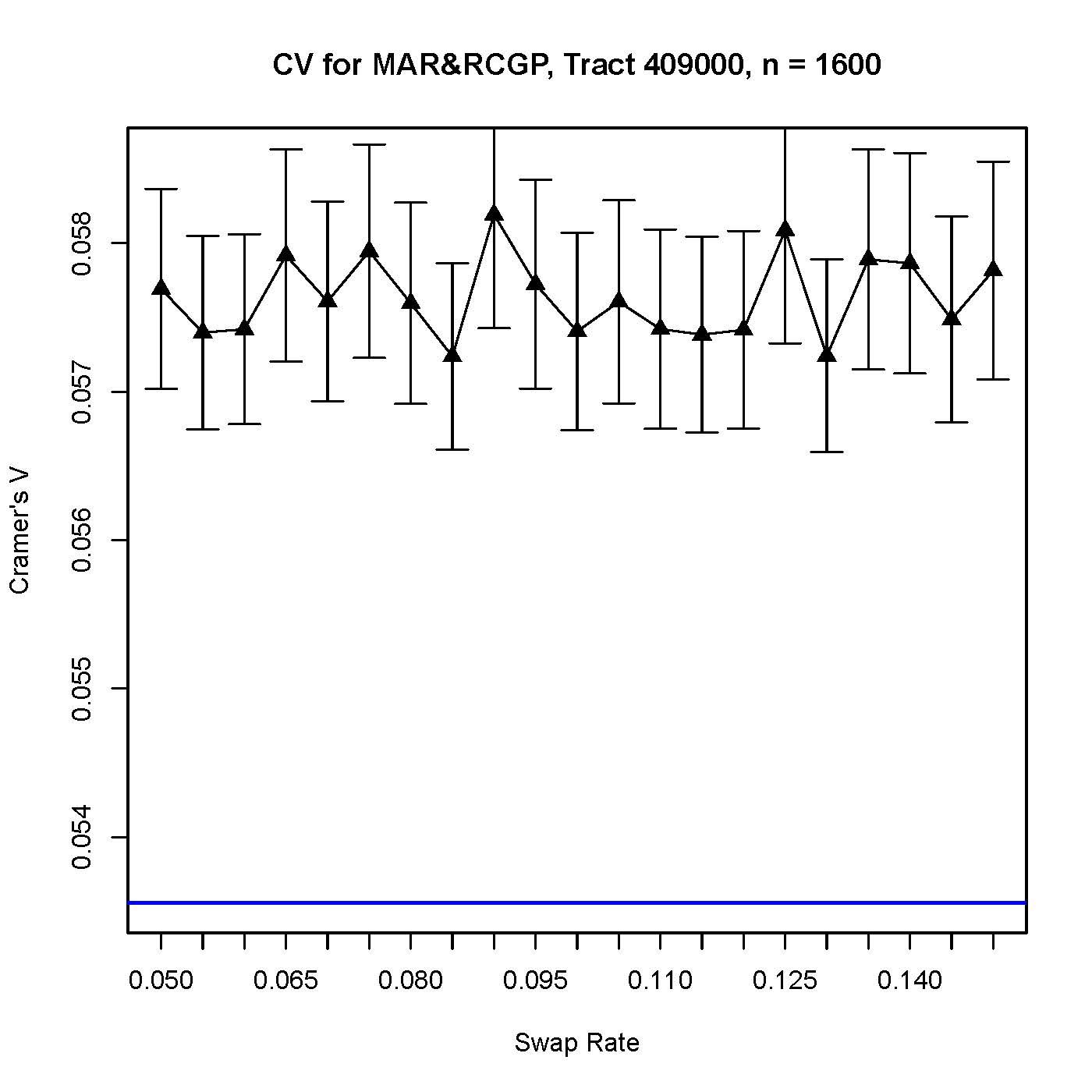}
\includegraphics[width=0.49\textwidth]{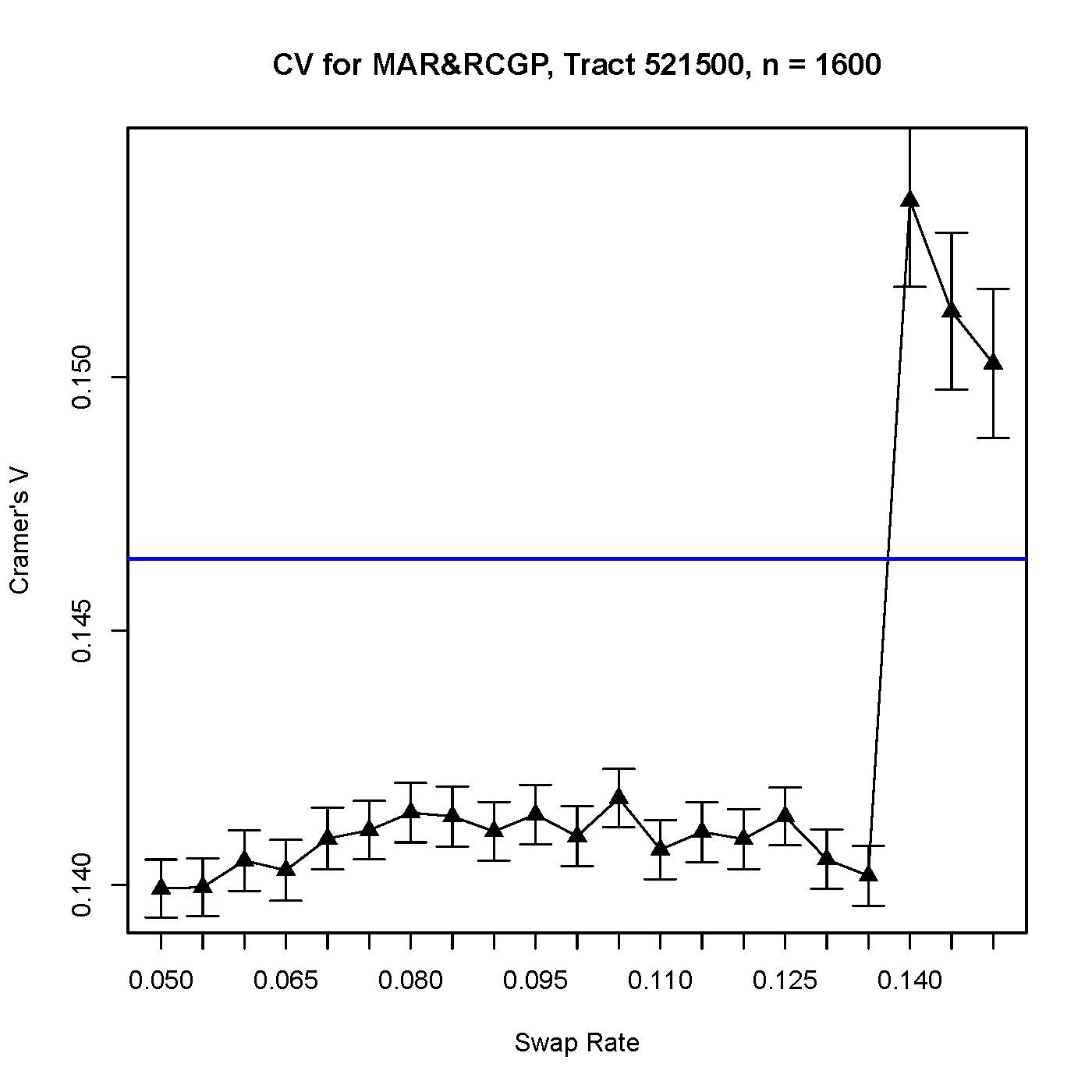}
\caption{ Cram\'er's $V$ versus swap rates for the tables of marital status by
race (not sparse, but weaker association). }
\label{fig:unswapplots2} 
\end{figure}

Targeted swapping was performed on these data. Figure \ref{fig:unswapplots}
(simulations on martial status by age) and Figure \ref{fig:unswapplots2}
(marital status by race) demonstrates the amount of variability in the behavior
of the Cram\'er's $V$ values as the swap rate is increased. The instability in
the joint distributions is observable in both the data with synthetic joint
distributions and the data with the true sampled joint distributions. Each of
the six different tracts shown in the plot titles is the largest (by population)
tract for six different PUMAs in Allegheny County, out of its nine PUMAs. 1,600
simulated swaps were performed at each swap rate between 5\% and 15\% in
increments of 0.5\%. 

\section{Discussion}
Data swapping in its simplest form, wherein a fraction of households is swapped
at random, will ``normalize'' the strengths of the joint distributions of
categorical variables, instead of lowering them. This effect is still observed
even when a primitive matching stage is included, so that two households may
only be swapped if they match on some predefined set of key variables. However,
the further addition of a minimal targeting stage in the data swapping procedure
is shown to impact the statistical quality of the data in an inconsistent way:
by deciding to implement a generic selection criterium for at-risk households,
even the expected direction of swapping's effect on the joint distributions can
no longer be predicted. 

The goal of this simulation study was to understand the impact of data swapping
on statistical analyses of U.S.\ Census data. One way this work only
approximates this goal is that the specifics of this data swapping algorithm
cannot perfectly match what the Census Bureau has, since the true details of
that algorithm are hidden from the public. Inspiration was drawn from what is
publicly known about the true data swapping method, but these results are only a
step towards the right direction. The targeting stage of the algorithm would be
a natural focus for a future analysis; it may be that a different choice of
at-risk variables used for selecting the set of records to swap may lessen the
effect noticed in this paper, at least for a particular set of variables whose
joint distributions are of particular interest. 

Another point of interest is in expanding the list of statistical measures
considered their susceptibility to data swapping's effects. Cram\'er's $V$ was
merely chosen due to its similarity to the ubiquitous Pearson's chi-square, but
many other measures and statistical procedures are of interest to researchers
across fields. In particular, an analysis of more robust measures of association
than Pearson's chi-square (and by extension, Cram\'er's $V$) would be of use in
separating the effect of data swapping and data sparsity. In this analysis,
collapsing variables served as a stand-in for a more robust measure whenever the
data were deemed to be too sparse. A follow-up study may better capture the true
effects of data swapping by considering a non-asymptotic statistic.  

The $n$-Cycle swap procedure is based on the idea of data swapping, except
groups of $n$ individuals are chosen simultaneously and their records are
permuted. In other words, data swapping is $n$-Cycling in the case where $n =
2$; the procedure and its benefits are explained in detail in an article by
\citet{DePersio:wd}. Some extension of the approach outlined in this paper may
be useful for understanding the effect of $n$-Cycling on statistical measures.
The process of evaluating the distortions to statistical analyses should be
updated in this time when researchers have ever-increasing access to public
census data and computational resources.

\section*{Acknowledgements}
William F.\ Eddy and Stephen E.\ Fienberg served as advisors to this work at all
stages, not only with the technical aspects, but also with the process of
obtaining and understanding the data. I would like to thank Laura McKenna,
Marlow Lemons, Amy Lauger, Michael Freiman, and Paul Massell for helpful
discussions about the U.S.\ Census Bureau's own data swapping implementation and
also for their efforts in making my research visits to the Census Bureau as
productive as possible. The Carnegie Mellon NSF-Census Research Node provided
feedback at nearly all stages of this work. I am also grateful to Rebecca C.\
Steorts and Peter Sadosky for invaluable discussions about data swapping and the
data. Finally, I thank the anonymous reviewer for providing numerous insightful
comments which led to a much clearer document. This work was partially supported
by NSF grant SES 1130706.

\bibliographystyle{plainnat}
\bibliography{swappaper}

\end{document}